# Fine Structure and Lifetime of Dark Excitons in Transition Metal Dichalcogenide Monolayers

C. Robert[1], T. Amand[1], F. Cadiz[1], D. Lagarde[1], E. Courtade[1], M. Manca[1], T. Taniguchi[2], K. Watanabe[2], B. Urbaszek[1], and X. Marie[1]

[1]*Université de Toulouse, INSA-CNRS-UPS, LPCNO, 135 Avenue de Rangueil, 31077, Toulouse, France*
[2]*National Institute for Materials Science, Tsukuba, Ibaraki 305-0044, Japan*

*The intricate interplay between optically dark and bright excitons governs the light-matter interaction in transition metal dichalcogenide monolayers. We have performed a detailed investigation of the "spin-forbidden" dark excitons in WSe2 monolayers by optical spectroscopy in an out-of-plane magnetic field $B_z$. In agreement with the theoretical predictions deduced from group theory analysis, magneto-photoluminescence experiments reveal a zero field splitting $\delta=0.6 \pm 0.1$ meV between two dark exciton states. The low energy state being strictly dipole forbidden (perfectly dark) at $B_z=0$ while the upper state is partially coupled to light with z polarization («grey» exciton). The first determination of the dark neutral exciton lifetime $\tau^D$ in a transition metal dichalcogenide monolayer is obtained by time-resolved photoluminescence. We measure $\tau^D \sim 110 \pm 10$ ps for the grey exciton state, i.e. two orders of magnitude longer than the radiative lifetime of the bright neutral exciton at T=12 K.*

## I. INTRODUCTION

Transition metal dichalcogenide (TMD) monolayers (MLs) are direct semiconductors with an energy gap in the visible region of the optical spectrum situated at the K-point of the Brillouin zone. They are an ideal platform to study light-matter interaction and spin-valley physics in the limit of two-dimensional carrier confinement[1,2,3,4]. The optical properties are governed by very robust excitons (Coulomb bound electron-hole pairs), with binding energies of the order of 500 meV [5,6,7,8,9,10,11,12]. These atomically thin materials show light absorption of up to 15 % per monolayer at exciton resonance energies. Not just the absorption strength is remarkable, but also the polarization properties. The lack of crystal inversion symmetry in TMD MLs together with the strong spin-orbit interaction in these materials leads to a coupling of carrier spin and k- space valley dynamics[13]. As a result, the circular polarization (σ+ or σ−) of the absorbed or emitted photon can be directly associated with selective exciton generation in one of the two non-equivalent K-valleys: K+ or K−, respectively[14,15,16,17].

To efficiently couple to light, the exciton transition has to be electric-dipole-allowed and spin-allowed *i.e.* it depends on the relative orientation of the electron and hole spins. An exciton for which the electron and the hole spins are oriented (anti-) parallel to each other is called a dark (bright) exciton. Dark excitons have orders of magnitude longer recombination times than bright excitons, usually determined by a tiny but essential coupling to optically active states. These long lived excitons in semiconductor nano-structures are to a certain extend decoupled from their environment and have been successfully used to investigate Bose-Einstein condensation[18,19] and to implement quantum information protocols[20,21,22]. It is therefore crucial to understand how to optically initialize dark excitons, their energy fine structure and coupling to light.

The key role played by these neutral dark excitons on the optical properties of TMD ML has been recently demonstrated in various photoluminescence (PL) or electroluminescence spectroscopy experiments[23,24,25,26]. In particular the *increase* of the luminescence intensity in WSe$_2$ monolayers when the temperature increases is the consequence of the interplay between bright and dark exciton populations. These results are in agreement with ab-initio calculations of the bright-dark exciton splitting energy which predict that in WX$_2$ systems, dark states are lower in energy than bright ones, whereas this order is reversed for MoX$_2$ MLs [27,28,29,30]. This is mainly due to the change of sign of the spin-orbit splitting in the conduction band between these different materials. The exact amplitude of the bright-dark exciton splitting $\Delta$ depends both on the conduction band spin-orbit splitting and the exchange interaction between electron and hole [27].

This bright-dark exciton splitting $\Delta$ has been measured very recently in WSe$_2$ monolayers encapsulated in hBN by exploiting the optical selection rules associated to in-plane propagation of light in the monolayer[31]. An energy $\Delta = 40 \pm 1$ meV has been determined in excellent agreement with another measurement technique based on near-field coupling to surface plasmon polaritons[32]. Slightly larger values $\Delta \sim 47$ meV were determined in non-encapsulated WSe$_2$ monolayer with experiments mixing the bright and dark excitons under high transverse magnetic fields (parallel to the ML plane)[33,34]. This larger value is consistent with a larger exchange interaction due to the expected smaller dielectric screening.

It was recently predicted that the short-range exchange interaction between the electron and the hole should also lift the double degeneracy of dark neutral excitons[35,36]. However, an experimental evidence of such a fine structure splitting of dark exciton in TMD is still lacking. Such splitting due to exchange interaction of non-optically active *direct* exciton was also predicted in other 2D semiconductor structures (III-V or II-VI semiconductor quantum wells[37,38]) while it has never been measured (to the best of our knowledge) probably because of its extremely small value.

In this paper we present magneto-photoluminescence measurements in a magnetic field B$_z$ applied perpendicular to the ML plane, which allow us to measure the dark exciton energy splitting $\delta$. We find $\delta$=0.6 $\pm$ 0.1 meV in WSe$_2$ ML between the lowest energy state which is perfectly dark at B$_z$=0 and the upper state which is partially coupled to light in z polarization (« grey » exciton).

Moreover we measure for the first time the lifetime of dark neutral exciton in TMD ML[39]. We find that it is two orders of magnitude longer than the one of bright excitons at cryogenic temperature, typically 110 ps in WSe$_2$ monolayers. This is in agreement with recent theoretical predictions[35].

The paper is organized as follows. The next section presents the symmetry analysis of the exciton fine structure. In section III, we present the sample fabrication and the experimental setups. Magneto-PL measurements are presented in section IV while time-resolved PL experiments are shown in section V.

## II. FINE STRUCTURE OF DARK EXCITONS FROM SYMMETRY ANALYSIS

In this section we give details on the two types of spin-forbidden exciton states. The lower energy state is not electric-dipole active (perfectly dark), whereas the higher energy state can couple to light with an out of plane polarization z-mode (grey exciton).

The point symmetry group of a TMD ML is D$_{3h}$. Since the direct band gap is localized at the edges of hexagonal Brillouin zone, the symmetry of individual valley K$\pm$ is lower and is described by the C$_{3h}$ point group. The symmetry of the electronic states of the two valleys at K points can be derived from merging the two valleys equivalent wave-vector groups ($C_{3h}$) corresponding to each valley and using the compatibility tables of $D_{3h}$ and $C_{3h}$ [31,40]. As the single particle Hamiltonian commutes with any operations of the crystal point group, it is possible to choose a complete set of electron eigen-states in the irreducible representations of $D_{3h}$ at $K_\tau$ ($\tau = \pm 1$) points. Figure 1a shows the obtained single particle states at these high symmetry points. In contrast to the more common use of the representation of $C_{3h}$, we shall see that this approach is more practical when dealing with excitons. The obtained electronic states are denoted here as: $U_{v,\tau,s}^{\gamma}(\mathbf{r}) \equiv e^{i\mathbf{K}_\tau \cdot \mathbf{r}} u_{v,\mathbf{K}_\tau,s}^{\gamma}(\mathbf{r})$, where $u_{v,\mathbf{K}_\tau,s}^{\gamma}(\mathbf{r})$ is the periodic part of the Bloch function, $\mathbf{K}_\tau$ is the in-plane valley wave-vector, $\gamma$ is the irreducible representation (irrep) index in notations of Ref. [40], and $s = \pm 1/2$ is the *effective* spin index labelling the two states in the 2-dimension irrep $\Gamma_\gamma$ [40]. The two valleys correspond each other by time-reversal, so that we have in general: $\hat{K}\left[U_{v,\tau,s}^{\gamma}(\mathbf{r})\right] = (-1)^{1/2-s}\left(U_{v,-\tau,-s}^{\gamma}(\mathbf{r})\right)^*$, where $\hat{K}$ is the Kramers time-reversal operator.

Note that this labelling implicitly takes into account spin-orbit mixing. For instance, $U_{v,+1,1/2}^{7}(\mathbf{r})$ is mixed with $U_{v-1,+1,-1/2}^{7}(\mathbf{r})$, so that besides a main spin contribution $\uparrow$, it contains a small component of $\downarrow$ spin state due to mixing with the band $v-1$ states [31].

Here we restrict our description to A excitons composed of an electron from one of the two conduction bands split by the spin-orbit interaction $\Delta_{SO}$ and a hole from the upper valence band A (figure 1a).

(i) *Bright excitons $X^0$ of $\Gamma_6$ representation built with $\Gamma_9$ conduction electrons*:

For $\Gamma_9$ conduction electron states, and the highest $\Gamma_7$ valence band, the exciton states belong to the reducible representation:

$$\Gamma_{7,v}^{9,c} = \Gamma_9^c \otimes \Gamma_7^h = \Gamma_9^c \otimes \Gamma_7^{v*} = \Gamma_9^c \otimes \Gamma_7^v$$

$$i.e. \quad \Gamma_{7,v}^{9,c} = \Gamma_5 \oplus \Gamma_6 \tag{1}$$

As a consequence, the 4-dimensions exciton representation is the direct sum of two independent two-dimensional irreducible representations. Using now the coupling tables of $D_{3h}$ group, the $\Gamma_6$ bright exciton Bloch functions are derived below.

$$\Psi_{-1}^{6}(\mathbf{r}_e, \mathbf{r}_h) = U_{-1,+1/2}^{9,c}(\mathbf{r}_e)U_{+1,+1/2}^{7,h}(\mathbf{r}_h) = -U_{-1,+1/2}^{9,c}(\mathbf{r}_e)\hat{K}\left(U_{-1,-1/2}^{7,v}(\mathbf{r}_h)\right)$$

$$\Psi_{+1}^{6}(\mathbf{r}_e, \mathbf{r}_h) = U_{+1,-1/2}^{9,c}(\mathbf{r}_e)U_{-1,-1/2}^{7,h}(\mathbf{r}_h) = U_{+1,-1/2}^{9,c}(\mathbf{r}_e)\hat{K}\left(U_{+1,1/2}^{7,v}(\mathbf{r}_h)\right) \tag{2}$$

They correspond to two degenerate *intra-valley* excitons ($\tau_c = \tau_v$), which transform like in-plane vectors under the symmetry operations of $D_{3h}$. Note that their main spin components are identical in both electron and valence states. These states, are located at the $\Gamma$ point of the exciton Brillouin zone and are optically active (bright excitons)[41].

In the same way we can show that the two excitons with $\Gamma_5$ representation are optically forbidden and are located at the $K_\tau$ point of the exciton Brillouin zone (k-space indirect and spin forbidden transitions).

(ii) *Dark excitons $X^D$ of $\Gamma_3$ and $\Gamma_4$ representations:*

For $\Gamma_8$ conduction electron states, and the highest $\Gamma_7$ valence band, the exciton states belong to the reducible representation:

$$\Gamma_{7,v}^{8,c} = \Gamma_8^c \otimes \Gamma_7^h = \Gamma_8^c \otimes \Gamma_7^{v*} = \Gamma_8^c \otimes \Gamma_7^v$$

(3)

$$i.e. \quad \Gamma_{7,v}^{8,c} = \Gamma_3 \oplus \Gamma_4 \oplus \Gamma_6$$

In this case the 4-dimensions exciton representation is the direct sum of two independent one-dimensional irreducible representations ($\Gamma_3$ and $\Gamma_4$) and one two-dimensional irreducible representation ($\Gamma_6$). Using the coupling tables of D$_{3h}$ group, we can show that the two excitons with $\Gamma_6$ representation are spin allowed but are indirect in k-space. This paper focuses on the two other direct excitons with $\Gamma_3$ and $\Gamma_4$ representations whose Bloch functions are derived below.

$$\Psi^3 = \frac{1}{\sqrt{2}}\left(U_{+1,+1/2}^{8,c} U_{-1,-1/2}^{7,h} - U_{-1,-1/2}^{8,c} U_{+1,+1/2}^{7,h}\right) = \frac{1}{\sqrt{2}}\left[U_{+1,+1/2}^{8,c} \hat{K}\left(U_{+1,+1/2}^{7,v}\right) + U_{-1,-1/2}^{8,c} \hat{K}\left(U_{-1,-1/2}^{7,v}\right)\right]$$

(4)

$$\Psi^4 = \frac{i}{\sqrt{2}}\left(U_{+1,+1/2}^{8,c} U_{-1,-1/2}^{7,h} + U_{-1,-1/2}^{8,c} U_{+1,+1/2}^{7,h}\right) = \frac{i}{\sqrt{2}}\left[U_{+1,+1/2}^{8,c} \hat{K}\left(U_{+1,+1/2}^{7,v}\right) - U_{-1,-1/2}^{8,c} \hat{K}\left(U_{-1,-1/2}^{7,v}\right)\right]$$

Remarkably these exciton states correspond to a coherent superposition of *intra-valley* conduction-valence pairs, the only difference lying in the relative phase between the pairs ($\pm 1$). The $\Gamma_3$ exciton transforms like a pseudo-scalar, while the $\Gamma_4$ like the z-component of a vector in D$_{3h}$. These states lay at the $\Gamma$ point of the exciton Brillouin zone, but only the $\Gamma_4$ exciton is optically active (for z-polarized modes)[31,35]. It is this dark state that has been identified very recently in the optical spectroscopy experiments performed in zero external magnetic fields[31,32,42]. In contrast the $\Gamma_3$ exciton is truly dark, as can be demonstrated from dipolar selection rules. The optical transitions properties of all these direct excitons are summarized in table I.

As it was shown previously, the energy difference $\Delta$ between the dark excitons X$^D$ ($\Gamma_3$, $\Gamma_4$) and the bright excitons X$^0$ ($\Gamma_6$) depends both on the spin-orbit splitting in the conduction band $\Delta_{SO}$ and the short range part of the electron-hole exchange interaction[27]. It turns out that the dark excitons states $\Gamma_3$ and $\Gamma_4$ are also separated from each other with a splitting $\delta$ due to the short range exchange interaction, as shown below.

From the theory of invariants, it can be easily deduced that the general form of the short range electron-hole exchange Hamiltonian takes the form :

$$\hat{H}_{exch}^\gamma = 2a_\gamma \hat{S}_z \hat{S}_z^h - \frac{b_\gamma}{2}\left(\hat{S}_+ \hat{S}_-^h + \hat{S}_- \hat{S}_+^h\right)$$

(5)

where $\hat{S}_z$ ( $\hat{S}_z^h$ ) and $\hat{S}_\pm = \mp\left(S_x \pm iS_y\right)$ ( $\hat{S}_\pm^h = \mp\left(\hat{S}_x^h \pm i\hat{S}_y^h\right)$ ) are conduction (valence) electron effective spin operators, which belong to $\Gamma_2$ and $\Gamma_5$ representations of D$_{3h}$ respectively, and $\gamma$ is the exciton representation index (here, $\gamma = 3,4,5,6$). Let us determine the matrix representation of $\hat{H}_{exch}$ in the subspace $\left\{Y^3, Y^4, Y_{+1}^6, Y_{-1}^6\right\}$, restricted here to the 1s A direct exciton states.

In the $\left\{ \mathsf{Y}_{+1}^6, \mathsf{Y}_{-1}^6 \right\}$ bright exciton subspace, the exchange Hamiltonian reduces simply to $\hat{H}_{exch}^6 = \dfrac{a_6}{2} I$ , where $I$ is the 2×2 identity matrix[43].

In the reducible representation $\Gamma_3 + \Gamma_4$ generated by $\left\{ \mathsf{Y}^4, \mathsf{Y}^3 \right\}$ dark exciton states, the exchange Hamiltonian can be *a priori* written as:

$$\hat{H}_{exch}^{34} = \frac{1}{2} \begin{pmatrix} -a_4 + b_4 & 0 \\ 0 & -a_3 - b_3 \end{pmatrix} \tag{6}$$

, where $a_{3(4)}$ correspond to the first term of the Hamiltonian (5) and $b_{3(4)}$ comes from the swapping between the two direct electron-hole pairs entering into the linear combination which constitute $\mathsf{Y}^4$ and $\mathsf{Y}^3$ dark states. Since the orbital parts in the two valleys are conjugate, we can thus infer that: $a_3 = a_4 = a_d$ and $b_3 = b_4 = \delta$ , so that we get the simpler form:

$$\hat{H}_{exch}^{34} = \frac{1}{2} \begin{pmatrix} -a_d + \delta & 0 \\ 0 & -a_d - \delta \end{pmatrix} \tag{7}$$

This expression shows clearly that $\mathsf{Y}^4$ and $\mathsf{Y}^3$ states are split by exchange[35,36]. Finally, neglecting the mixing between direct and indirect $\Gamma_6$ excitons[43], we can further approximate: $a \equiv a_d = a_6$.

Including spin-orbit interaction, we get the final form in the subspace $\left\{ \mathsf{Y}^3, \mathsf{Y}^4, \mathsf{Y}_{+1}^6, \mathsf{Y}_{-1}^6 \right\}$:

$$\hat{H}_{exch} + \hat{H}_{so} = \begin{pmatrix} E_0 - \dfrac{a}{2} - \dfrac{\delta}{2} & 0 & 0 & 0 \\ 0 & E_0 - \dfrac{a}{2} + \dfrac{\delta}{2} & 0 & 0 \\ 0 & 0 & E_0 + \Delta_{so} + \dfrac{a}{2} & 0 \\ 0 & 0 & 0 & E_0 + \Delta_{so} + \dfrac{a}{2} \end{pmatrix} \tag{8}$$

where $\Delta_{so}$ is the conduction spin-orbit splitting, and $E_0$ is the 1s exciton energy.

The measured splitting between the bright and grey excitons is $\Delta = \left( \Delta_{so} + a - \dfrac{\delta}{2} \right)$ , whereas the fine structure splitting between the grey and "truly" dark exciton is $\delta$ .

Figure 1b summarizes the exciton fine structure at the $\Gamma$ point of the Brillouin zone with arrows corresponding to the bright $X^0$ and dark $X^D$ exciton transitions.

*Coupling of the dark excitons by a longitudinal magnetic field $B_z$.*

From group symmetry compatibility tables we can show that the grey ($\Gamma_4$) and truly dark excitons ($\Gamma_3$) can couple to each other in an applied magnetic field $B_z$. The corresponding Hamiltonian in $\Gamma_3 + \Gamma_4$ can be written as :

$$\hat{H} = -\frac{\delta}{2}\hat{\sigma}_z + \frac{g^D \mu_B B_z}{2}\hat{\sigma}_y \qquad (9)$$

where $\sigma$ are formal Pauli matrices written in the basis ($\Psi^3$, $\Psi^4$).

In a longitudinal magnetic field, both exciton states become grey (they can couple to light). Their energy and eigen-states are:

$$\begin{cases} \lambda_- = E_0 - \frac{1}{2}\sqrt{\delta^2 + \left(g^D \mu_B B_z\right)^2} \quad ; \qquad \left|\Psi^-\right\rangle = \cos\frac{\theta}{2}\left|\Psi^3\right\rangle - i\sin\frac{\theta}{2}\left|\Psi^4\right\rangle \\[2mm] \lambda_+ = E_0 + \frac{1}{2}\sqrt{\delta^2 + \left(g^D \mu_B B_z\right)^2} \quad ; \qquad \left|\Psi^+\right\rangle = \sin\frac{\theta}{2}\left|\Psi^3\right\rangle + i\cos\frac{\theta}{2}\left|\Psi^4\right\rangle \end{cases} \qquad (10)$$

where $\theta \in \left]-\pi/2, +\pi/2\right[$ is defined by :

$$\begin{cases} \cos\theta \equiv \dfrac{\delta}{\sqrt{\delta^2 + \left(g^D \mu_B B_z\right)^2}} > 0 \\[4mm] \sin\theta \equiv \dfrac{g^D \mu_B B_z}{\sqrt{\delta^2 + \left(g^D \mu_B B_z\right)^2}} \end{cases} \qquad (11)$$

For very large magnetic fields, we thus get two fully mixed grey states with z-polarization.

On this basis, one can easily show that the oscillator strengths $f$ of the two "dark" excitons vary as a function of the magnetic field according to :

$$f_{\Psi_+}(B_z) = f_{\Psi_+}(0)\cos^2\frac{\theta}{2} = \frac{1+\cos\theta}{2} \quad \text{and} \quad f_{\Psi_-}(B_z) = f_{\Psi_+}(0)\sin^2\frac{\theta}{2} = \frac{1-\cos\theta}{2} \qquad (12)$$

where $\Psi^+$ ($\Psi^-$) is the upper (lower) dark state and $f_{\Psi_+}(0)$ is the oscillator strength of the grey transition at zero magnetic field.

## III. SAMPLE AND EXPERIMENTAL SET-UP

High quality $WSe_2$ MLs encapsulated in hexagonal boron nitride (hBN) and transferred onto an $SiO_2$ (90 nm)/Si substrate have been fabricated. Details on sample fabrication can be found in ref.[44,45]. The typical thickness of the hBN layers is $\sim$ 10 nm and the in-plane size of the $MX_2$ ML is $\sim 10 \times 10$ $\mu m^2$. The encapsulation of $MX_2$ ML in atomically flat hBN layers reduces significantly the inhomogeneous broadening in the photoluminescence (PL) or reflectivity spectra, with typical values in the range 2−5 meV at low temperature[44,45,46,47]. These narrow exciton lines are crucial for the determination of the fine structure of the dark excitons presented in this paper.

Micro-PL experiments are performed in a standard geometry configuration in which the excitation and detection light propagates perpendicular to the ML plane. Nevertheless, by using high numerical aperture (NA) objectives, we have shown that part of the light propagating parallel to the ML can be collected[31]. This enables us to observe the z-polarized dark exciton transition. For technical reasons, different NA objectives have been used depending on the experiment setups. Their values are specified in the captions of each figure.

Experiments at T = 4 K and in longitudinal magnetic fields up to ±9 T have been carried out in an ultra-stable confocal microscope with a fiber coupling to the laser and detector system[48]. The detection spot diameter is about 700 nm. The sample is excited by a He- Ne laser (1.96 eV). The average laser power is in the µW range, in the linear absorption regime. The PL emission is dispersed in a double-monochromator and detected with a Si-CCD camera. The spectral resolution of this detection system is ~20 µeV. For time-resolved photoluminescence experiments, the flakes are excited by ~ 1.7 ps pulses generated by a tunable mode-locked Ti:Sa laser with a repetition rate of 80 MHz. For time-resolved experiments, unless it is mentioned, the excitation laser energy is 1.783 eV and the laser average power is 50 µW, *i.e.* quasi-resonant excitation conditions (~60 meV above the bright neutral exciton energy). The PL signal is dispersed by a spectrometer and detected by a Hamamatsu synchro-scan Streak Camera C5680 with a typical time-resolution of 2 ps. In all the experiments the excitation laser is linearly polarized.

## IV DARK EXCITON ENERGY SPLITTING DEDUCED FROM MAGNETO-PL MEASUREMENTS

First we present the experimental results for WSe$_2$ MLs at zero magnetic field. In figure 2, we observe in the PL spectrum at T=4K three sharp emission features. In agreement with previous reports[5,31], the high energy emission at 1.723 eV is attributed to the neutral bright exciton X$^0$ (symmetry $\Gamma_6$) recombination whereas the peak at 1.690 eV is often interpreted in terms of charged exciton (trion) emission. Note that this peak could also be a phonon-assisted transition line. Indeed, no signature in reflectivity spectra is observed at this energy (see ref. [45]) and the residual doping in hBN encapsulated WSe$_2$ ML is very small, as recently confirmed in charge-tunable encapsulated samples[49]. The third PL line at 1.682 eV was recently identified as the recombination of the so-called z-polarized dark exciton X$^D$ [31]. The optical selection rules recalled in section II dictate that this transition is normally forbidden for normal incidence conditions used in standard optical spectroscopy measurements (see table I). However we recently showed that its detection is possible in micro-PL experiments using a microscope objective with high numerical aperture NA (here NA= 0.82). In this case, the electric field vector at the focal tail has a significant component along the z-axis, which enables excitation/detection of the X$^D$ transition even at the normal incidence [31]. This will allow us to investigate the fine structure and the lifetime of the dark excitons without resorting to complex geometry configurations (edge detection) or coupling to surface plasmon-polariton.

Next we discuss the changes observed in the PL spectra when applying a longitudinal magnetic field B$_z$ perpendicular to the monolayer plane. Such magneto-PL or magneto-reflectivity experiments were performed for bright excitons X$^0$. They demonstrate the lifting of the valley degeneracy (valley Zeeman effect) with the magnetic field and yield the measurement of the bright exciton g factor[50,51,52,53,54,55,56,57,58,59]. In our hBN-WSe$_2$ ML-hBN sample, we measured (not shown) a bright exciton g factor g$^B$ =-4.25 ±0.01 as already observed by different groups [55,57,58]. In figure 3a, we present for the first time the magneto-PL spectra with the fine structure of dark excitons. The excitation laser is linearly-polarized and we present the right circularly-polarized PL spectra $\sigma^+$ as a function of the magnetic field from $B_z$=-9 to +9 Teslas (similar results (not shown) are obtained for $\sigma^-$ polarized detection, as expected). Interestingly two transitions with similar intensities are clearly visible at high magnetic field. This is different from the case of bright exciton where one transition is observed in $\sigma^+$ polarization while the other one is observed in $\sigma^-$ polarization. The Zeeman splitting between the two dark exciton states is clearly evidenced at large magnetic fields, with a typical energy separation between the two peaks up to 5 meV at $B_z$=+9 or − 9 T. For high magnetic field, the energy splitting between the two lines is simply equal to $g^D\mu_B B_z$, with $g^D$ being the g-factor of dark excitons and $\mu_B$ the Bohr magneton.

Remarkably we observe a non-linear dependence of the Zeeman splitting in the small magnetic field range -2 <$B_z$< 2 T. This is in perfect agreement with the theoretical prediction based on the existence of an exchange splitting $\delta$ between the dark states at $B_z$=0 (see section II). We have fitted the 19 spectra of figure 3a with equation (10) (see figure 3b). This yields an accurate determination of both the dark exciton exchange splitting $\delta$ and the dark exciton g factor. We find $\delta$=0.6 ± 0.1 meV and $|g^D|$ =9.4 ± 0.1. The results displayed in figure 3b demonstrate that the higher energy dark state (with symmetry $\Gamma_4$) is partially coupled to light, *i.e.* a "grey state" with z-polarization at $B_z$=0 whereas the lower energy state (with symmetry $\Gamma_3$) is optical inactive, *i.e.* a truly dark state[35] . The lower energy PL peak becomes clearly visible for $B_z$> ~ |2| T as a consequence of the magnetic field induced mixing between the two dark states : the lower energy state gains oscillator strength (equation (12) in section II). Note that the measured splitting of the dark exciton $\delta$=0.6 meV is much smaller than the value of ~10 meV that was predicted for TMD ML in ref [35]. However it is larger than the one measured in type II indirect excitons in GaAs/AlGaAs superlattices[60] ($\delta$=1.7 μeV[61]) or in InGaAs quantum dots ($\delta$=1 μeV [21]), despite the stronger quantum confinement for the latter. Note that a value of $\delta$ ~ 10 μeV was inferred from the magnetic field dependence of the PL circular polarization of direct excitons in type I GaAs quantum wells in ref.[38]. However the authors claimed that a contribution to this splitting could be due to departure of the quantum-well symmetry from the ideal $D_{2d}$ symmetry.

## V. LIFETIME OF DARK EXCITONS

In order to get more information on the dark exciton states, we have measured their lifetime by time-resolved photoluminescence spectroscopy. Figure 4 displays the normalized bright and dark exciton kinetics following a ps laser excitation at T=12 K. The bright exciton $X^0$ decay time is ~2 ps (limited by time resolution), corresponding to the intrinsic radiative exciton recombination time already measured by different groups[62,63,64]. Remarkably the dark exciton lifetime is two orders of magnitude longer: we measure $\tau^D$~110 ps. This gives a lower boundary for the radiative lifetime of the dark exciton as this measured decay time can be limited by other relaxation channels (non-radiative, toward the truly dark exciton intervalley excitons or localized states). Nevertheless, we note that this ratio is in good agreement with the recently calculated bright and dark exciton radiative decay rates [35]. In contrast to the assumption of ref. [33], our results show that the dark excitons have a non-negligible oscillator strength even at $B_z$=0, probably induced by spin-orbit mixing with higher energy conduction bands and lower energy valence bands[27,31] . The excitation laser energy used in figure 4 is 1.783 eV (quasi-resonant conditions) but we found similar dark exciton lifetimes for excitation energies resonant with the $X^0$ transition (1.723 eV), the 2s state (1.859 eV), the B exciton (2.158 eV) or above the free-carrier band gap (3.18 eV). Note that we confirmed the measurement of this dynamics in a second hBN-WSe$_2$ ML-hBN sample (not shown) which indicates that this decay time is an intrinsic feature. Nevertheless, we note that the linewidth of the dark exciton transition observed in Figure 2 and Figure 3a is significantly broader (~1 meV) than the homogeneous linewidth corresponding to a population decay of 110 ps. This is probably due to inhomogeneous contribution that is not completely cancelled by the hBN encapsulation.

Next we have investigated the temperature dependence of the dark exciton states. Figure 5a presents the dependence of the bright $X^0$ and dark $X^D$ exciton energy as a function of temperature, probed by cw PL spectroscopy. We observe clearly the red shift of both lines (for $X^0$, we measure the same dependence in reflectivity experiments, not shown). We emphasize that we did not observe in this temperature range any blueshift of the PL peaks, which could have been the fingerprint of a transient change from a localized bright/dark exciton regime to a free bright/dark exciton regime. The intensity of the $X^0$ transition increases with temperature while the $X^D$ transition quenches above 50 K. In figure 5b the dark exciton kinetics are displayed for temperatures in the range 12-90 K. For a lattice temperature T < 40 K, the measured PL decay time does not depend on temperature. This absence of variation of the PL decay time in the the temperature range 12–40 K could support an interpretation of the dark exciton lifetime $\tau^D$~110 ps at T=12 K controlled by radiative recombination processes and not by non-radiative channels.
For larger temperatures (T > 50 K), we observe that the dark exciton PL decay time shortens to become similar to the bright exciton PL decay time (~15 ps at 90K).
We emphasize that the measurements displayed in figures 4 and 5 are performed in the absence of any external magnetic field. As a consequence of the fine structure of the dark excitons revealed in section IV, we probe the higher energy dark state (*i.e.* the so-called grey state, with symmetry $\Gamma_4$).

**CONCLUSION**

We investigate the fine structure and the lifetime of so-called dark excitons in Transition Metal Dichalcogenide monolayers. This is made possible by working with high numerical aperture objectives in micro-PL experiments, which allows the detection of bright excitons with electric dipoles in the monolayer plane but also the spin-forbidden exciton transition with out of plane dipoles (z-polarization mode). The luminescence spectra recorded in longitudinal magnetic fields reveal a zero field splitting δ between a low energy state which is strictly optical inactive (truly dark state, dipole forbidden) and a higher energy state which has a small but non-negligible oscillator strength in z- polarization ("grey" state). This dark exciton energy splitting due to the Coulomb exchange interaction is about 500 times larger than the one reported previously in other III-V or II-VI semiconductor nanostructures. We measure a grey state lifetime of about ~110 ps, two orders of magnitude longer than the one of the bright exciton at T=12 K. Truly dark states are identified in our work and we show that their coupling to light can be controlled on-demand thanks to a mixing with grey exciton states in longitudinal magnetic fields. This opens interesting perspectives for investigations of Bose-Einstein exciton condensates and quantum information applications with transition metal dichalcogenide materials.

**ACKNOWLEDGMENTS**

We are grateful to Mikhail Glazov for enlightening discussions. We thank ANR MoS2ValleyControl, ITN Spin-NANO Marie Sklodowska-Curie grant agreement No 676108, ERC Grant No. 306719 , Programme Investissements d Avenir ANR-11-IDEX-0002-02, reference ANR-10- LABX-0037-NEXT, Laboratoire International Associe Grant No. ILNACS CNRS-Ioffe for financial support. X.M. also acknowledges the Institut Universitaire de France. K.W. and T.T. acknowledge support from the Elemental Strategy Initiative conducted by the MEXT, Japan and JSPS KAKENHI Grant Numbers JP26248061, JP15K21722 and JP25106006.

**FIGURE CAPTION**

**Figure 1** :
(a) Sketch of the single particle band structure of $WSe_2$ monolayer for both valleys K($\pm$). For convenience, a complete set of electron eigen-states in the irreducible representations of $D_{3h}$ has been chosen (b) Sketch of the exciton fine structure at the $\Gamma$ point of the exciton Brillouin zone (see text). $\uparrow\downarrow+\downarrow\uparrow$ and $\uparrow\downarrow-\downarrow\uparrow$ refer to the coherent superposition of the *intra-valley* conduction-valence pairs described in equation 4 (the arrows represent the electron spin components).

**Figure 2** :
PL spectrum of hBN/$WSe_2$ ML/hBN heterostructure at 4 K. The out-of-plane-polarized dark exciton $X^D$ is observed 40 meV below the in-plane polarized bright exciton $X^0$ thanks to high numerical aperture of the objective (NA=0.82).

**Figure 3** :
(a) PL of dark exciton as a function of longitudinal (perpendicular to the ML plane) magnetic field $B_z$ at 4 K. Spectra are vertically shifted for clarity. Grey dashed lines are guides to the eyes pointing the energy of the transitions. An aspheric lens with NA=0.68 is used for excitation and detection. Polarization configuration is linear excitation and circular detection. At $B_z=0$ T only the grey exciton line is observed ; (b) Energy of dark exciton states as a function of $B_z$ extracted from (a). Solid lines represent fits with equation (10): $\lambda_{\pm} = E_0 \pm \frac{1}{2}\sqrt{(\delta^2+(g^D\mu_B B_z)^2)}$.

**Figure 4** :
Time resolved photoluminescence of hBN/$WSe_2$ ML/hBN heterostructure at 12K showing short decay time (2 ps) for the in-plane polarized bright exciton $X^0$ close to the time resolution limit and long decay time (110 ps) for the out-of-plane polarized dark exciton $X^D$.

**Figure 5** :
(a) Contour plot of PL intensity as a function of temperature. $X^D$ signature quenches above 50 K while the intensity of bright exciton $X^0$ increases. (b) Temperature dependence of the $X^D$ photoluminescence dynamics. Solid lines represent monoexponential decays.

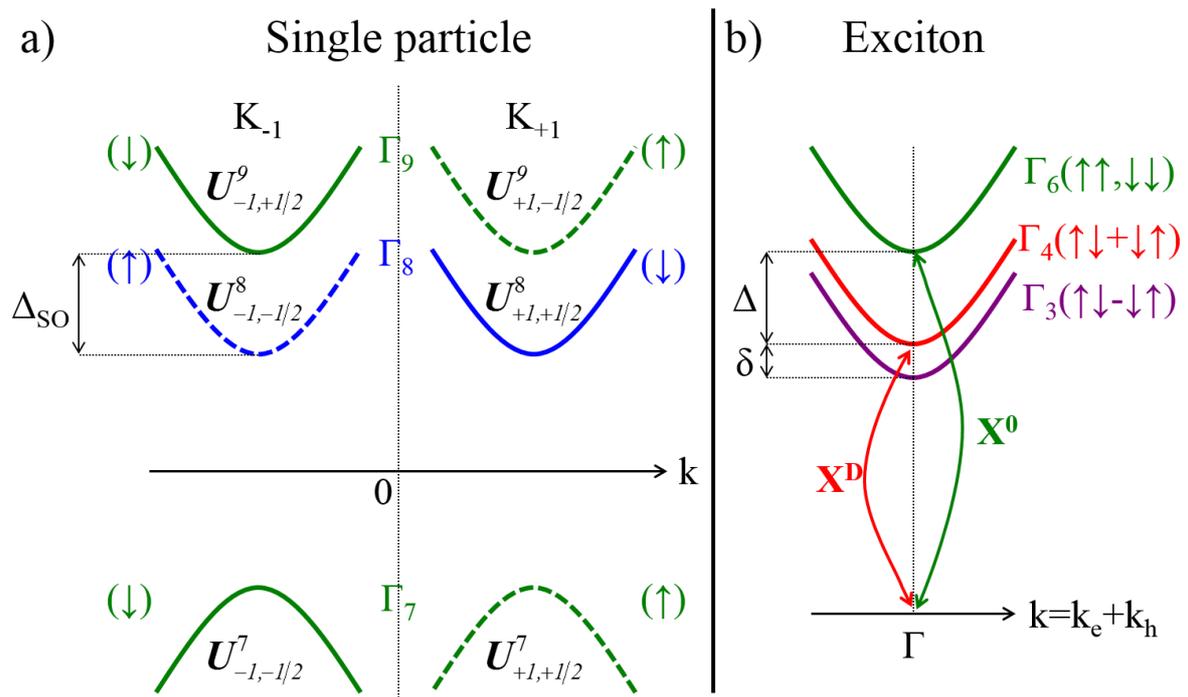

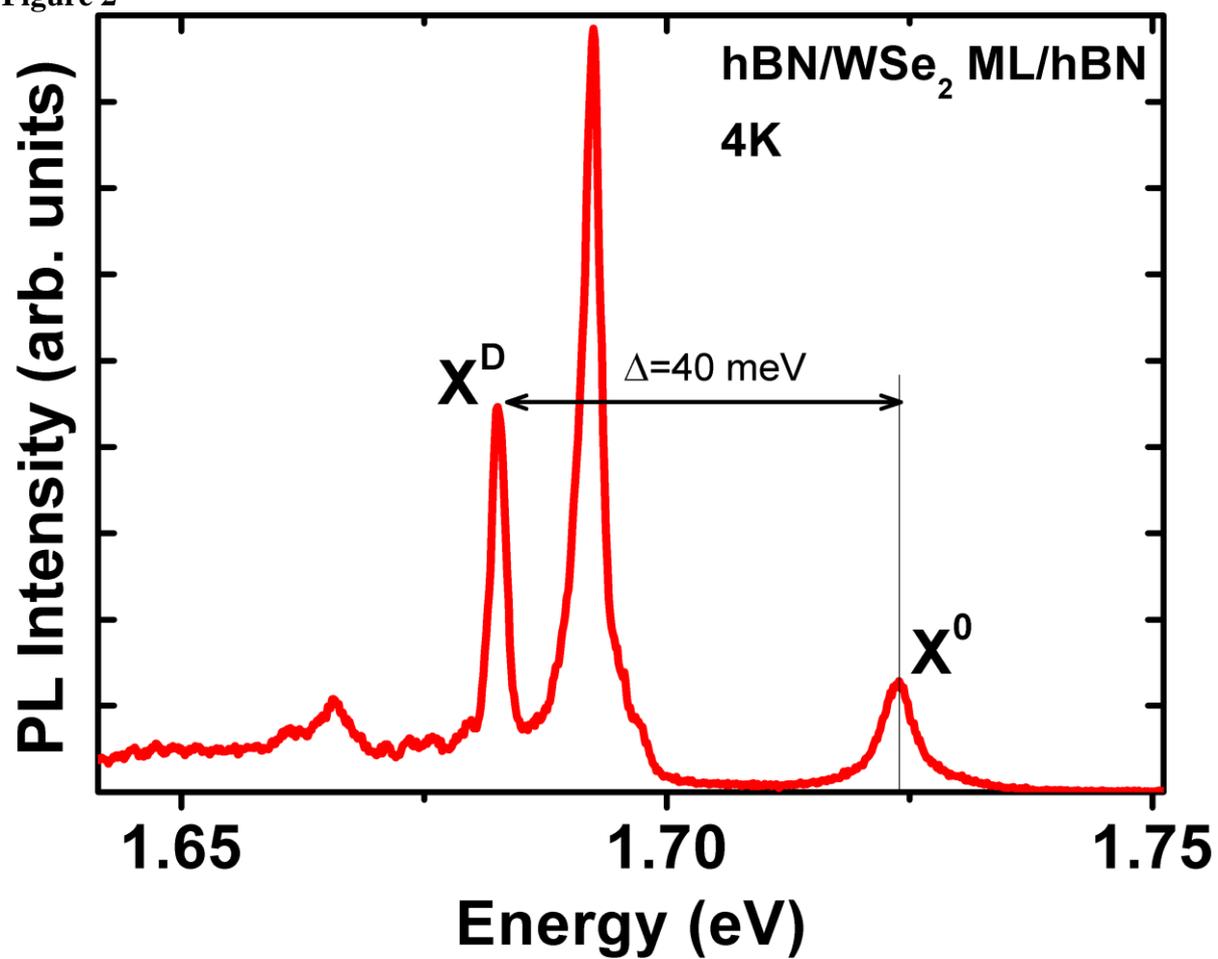

**Figure 3a**

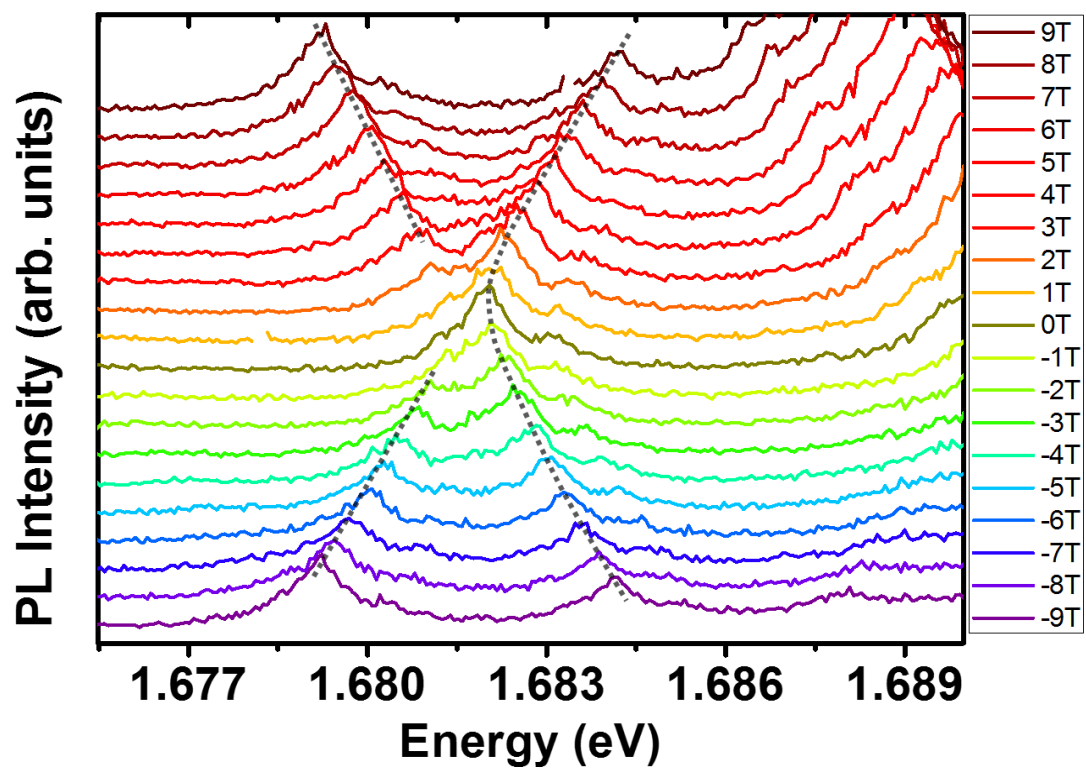

**Figure 3b**

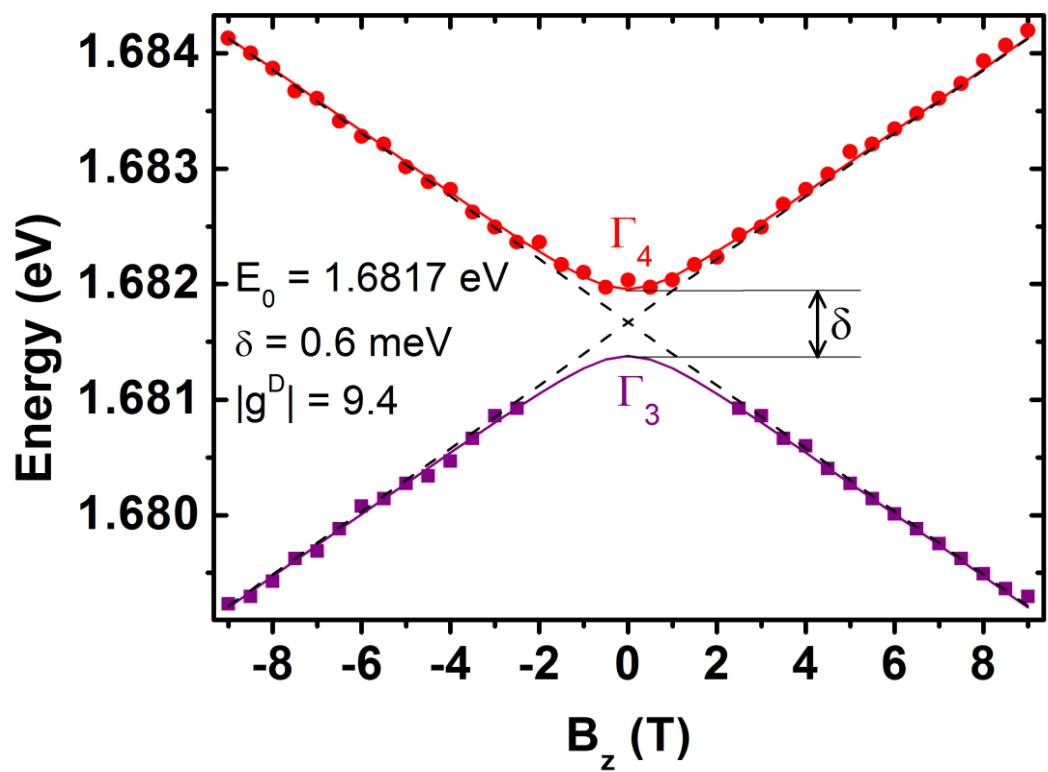

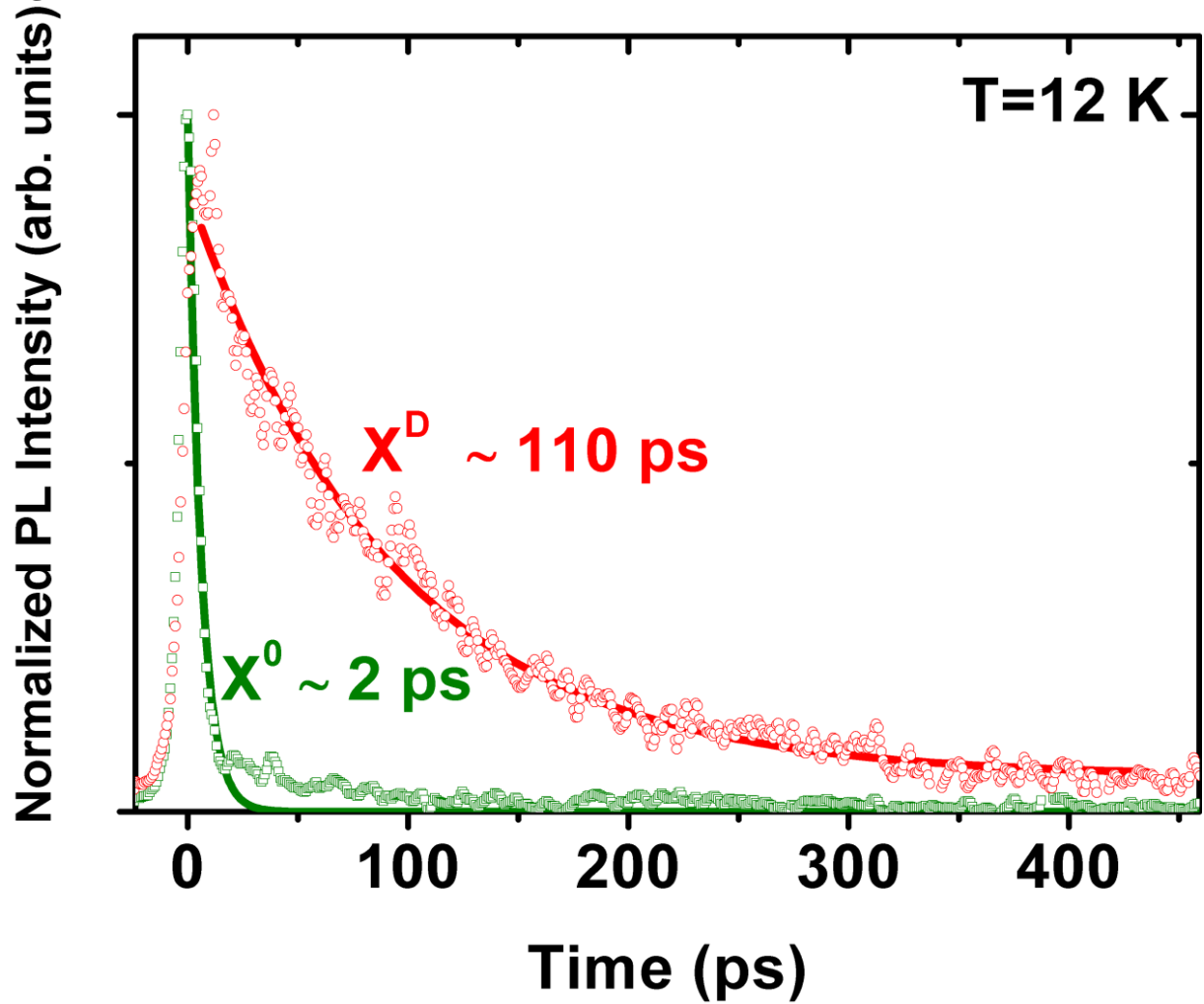

**Figure 5a**

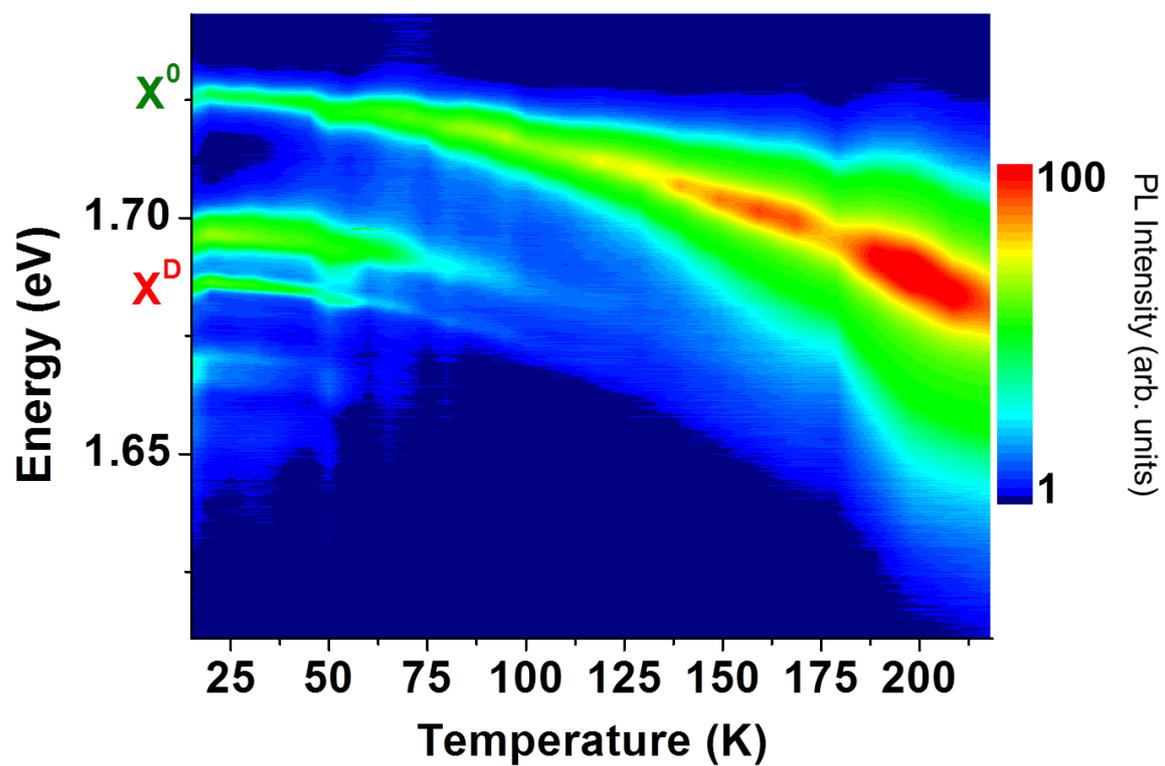

**Figure 5b**

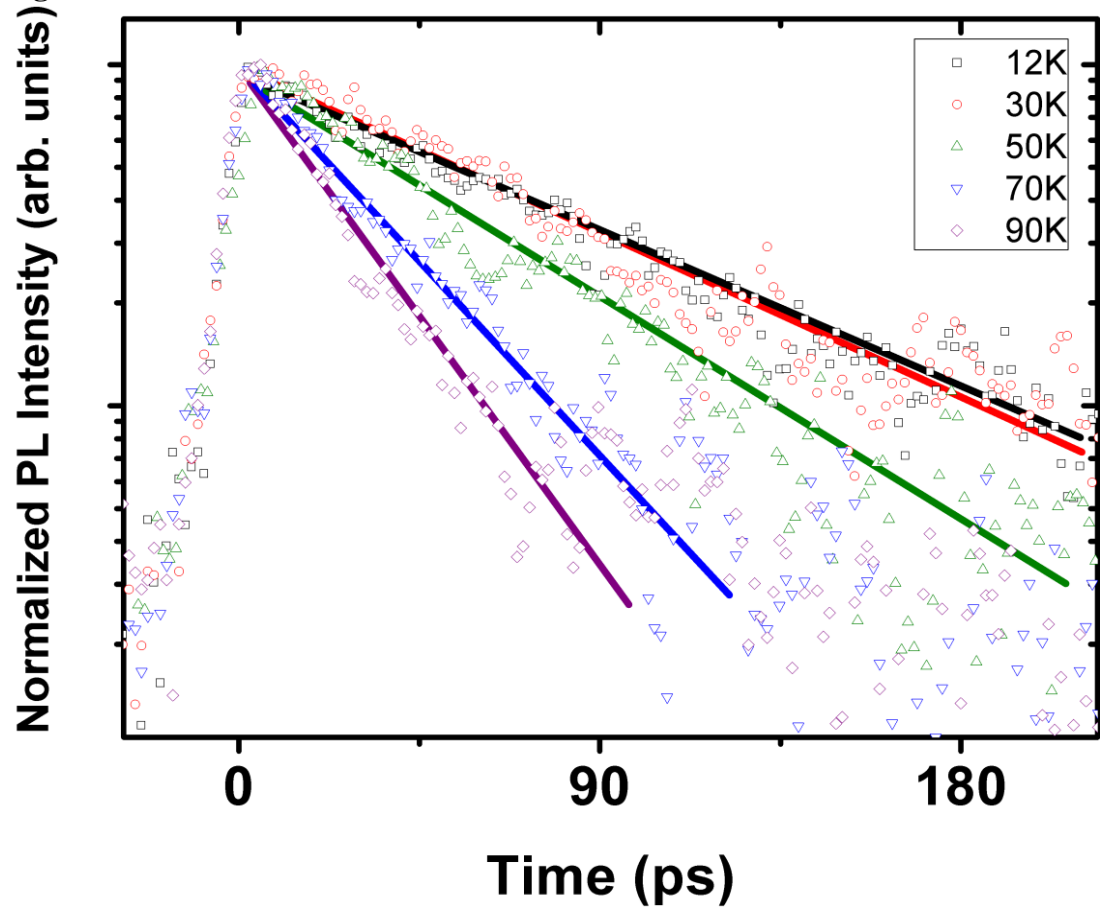

**Table I**

| Exciton states | Optical transitions for direct excitons | |
|---|---|---|
| $X^0$ bright, $\Gamma_6$ | Spin – allowed | Electric dipole allowed: dipole in ML plane |
| $X^D$ "dark" (grey), $\Gamma_4$ | "Spin – forbidden" | Electric dipole allowed: dipole out of ML plane (z-mode) |
| $X^D$ dark, $\Gamma_3$ | "Spin – forbidden" | Electric dipole forbidden |

**Table I Caption :**

The table gives an informal representation of the spin and dipole selection rules for the optical transitions studied[65]. The symmetry of the exciton states is taken from figure 1b, where the energetic ordering can be seen. It is important to note that the definition here of "Spin-forbidden" is approximate: For valence and conduction bands with absolutely *pure* spin states, both $\Gamma_3$ and $\Gamma_4$ dark excitons $X^D$ would be optical inactive (*i.e.* truly spin-forbidden). But recent PL experiments have revealed the recombination of $\Gamma_4$ excitons where the z-mode is detected[31,32]. This means a small spin mixing exists that makes this optical transition possible[31], hence the label "grey" indicating a small, but non-zero optical transition oscillator strength. The optical transition matrix elements from ground state towards each valley pairs with z-modes add up constructively for the grey $\Gamma_4$ states while they exactly cancel for the truly dark $\Gamma_3$ states.